# Magnetic and Hyperfine interactions in HoFe$_{1-x}$Cr$_x$O$_3$ ($0 \leq x \leq 1$) compounds

Ganesh Kotnana[1], V. Raghavendra Reddy[2] and S. Narayana Jammalamadaka[1*]

[1]*Magnetic Materials and Device Physics Laboratory, Department of Physics, Indian Institute of Technology Hyderabad, Hyderabad – 502 285, India.*

[2]*UGC-DAE Consortium for Scientific Research, Khandwa Road, Indore - 452 001, India.*

*Corresponding author: surya@iith.ac.in

**Abstract**

We report on the magnetic and Mössbauer properties of polycrystalline HoFe$_{1-x}$Cr$_x$O$_3$ ($0 \leq x \leq 1$) compounds. Magnetization data reveals the continuous tailoring of magnetic transition due to weakening of Ho$^{3+}$-Fe$^{3+}$ and Fe$^{3+}$-Fe$^{3+}$ interactions in the entire temperature range by replacing the Fe$^{3+}$ ions with Cr$^{3+}$ ions. The observed decrease in Néel temperature ($T_N$) and increase in spin re-orientation transition temperature ($T_{SR}$) with the replacement of Fe$^{3+}$ with Cr$^{3+}$ is ascribed to the weakening of Fe(Cr)-O-Fe(Cr) antiferromagnetic exchange interaction. In addition, we also attribute such a change in $T_N$ to the enhancement of ferromagnetic interaction of adjacent Cr$^{3+}$ moments through *t-e* hybridization as a result of the structural distortion. The decrease in isomer shift (IS) suggests enhancement of the interaction between nuclear charge with the 3*s* electrons as a result of decrease in radial part of 3*d* wave function with Cr addition. In this paper we also discuss about the variation of quadrupole splitting (QS) and hyperfine fields (H$_{hf}$) with Cr addition in HoFe$_{1-x}$Cr$_x$O$_3$ ($0 \leq x \leq 1$) compounds.

**Keywords:** Magnetization, Isomer shift, quadrupole splitting, ortho-ferrites, anti-ferromagnetism

**Introduction**:

Perovskites with general formula $ABO_3$ (A is rare earth ion or Yttrium and B is a transition metal) have been the most versatile compounds in oxide research. Compounds with variation of the simple perovskite structure exhibit interesting magnetic and electronic behaviour such as high temperature superconductivity [1], colossal magnetoresistance [2], half-metallicity [3] and multiferroism [4]. Rare earth ortho-ferrites ($RFeO_3$) have the complex spin structure between the rare earth (R) and the transition metal (TM) ions and are distorted perovskite family of canted antiferromagnets drawn considerable attention due to their unique physical properties [5] and potential applications such as ultrafast magneto-optical recording [6], laser-induced thermal spin re-orientation [7], precession excitation induced by terahertz pulses [8], inertia-driven spin switching [9], and magnetism-induced ferroelectric multiferroics [10]. Indeed the canted antiferromagnetism, Dzyaloshinskii – Moriya exchange interaction [11], very small anisotropy of Fe spins in a – c plane and large anisotropy along b – axis [12] may play vital role in determining their magnetic properties. The $Fe^{3+}$ magnetic moments in $RFeO_3$ undergo a phase transition into canted antiferromagnetic state exhibiting a weak ferromagnetic moment at a temperature $T_N$ ranging from 620 K to 750 K. One of the characteristic magnetic transition of $RFeO_3$ is spin re-orientation (SR) during which the direction of easy axis of magnetization of $Fe^{3+}$ sub lattice changes from one crystal axis to another upon varying the temperature [13]. The complex magnetic interaction between the two sub-lattices of $R^{3+}$ and $Fe^{3+}$ in ortho-ferrites makes them interest to study their magnetic properties.

Interestingly, $HoFeO_3$ is a representative of rare earth ortho-ferrites which has been believed to possess canted G-type antiferromagnetism and potential candidate for ultrafast recording [9] with magnetic ordering temperature around 641 K [14]. For example, $HoFeO_3$ exhibits the SR transition of $Fe^{3+}$ at 50-60 K with magnetic structure changing from $G_xA_yF_z$ ($\Gamma_4$) to

$F_xC_yG_z$ ($\Gamma_2$), where $G_x$, $A_y$, and $F_z$ stand for spin component along x, y, and z axis in terms of Bertaut's notation, respectively [15]. In recent years, it has been believed that variety of interesting properties can be achieved by alloying different kinds of cations at B-site of perovskite materials [16]. The only trivalent nature of elements in the structure of these materials make them attractive for isovalent substitutions. According to Goodenough-Kanamori theory, $Cr^{3+}$ is best choice for $Fe^{3+}$ to show superior magnetic property due to super exchange interaction [17]. Complex magnetic behaviour has been reported on $Nd(Cr_{1-x}Fe_x)O_3$ and $Y(Fe_{1-x}Cr_x)O_3$ respectively [18-19].

Studies on the $AB_xB'_{1-x}O_3$ perovskite compounds show that the structure and physical properties of these compounds strongly depends on the ionic size and charge differences of the B-site cations (B and B') [20]. As the physical properties of materials depend very sensitively on their structure, it is essential to know their structure which can be determined by the microscopic techniques such as X-ray diffraction. On the other hand, it is more important to know the local interactions of these compounds. Mössbauer spectroscopy [21-23] is an important microscopic technique to determine local interactions via hyperfine interaction, i.e. the interactions between the nuclear charge distribution and extranuclear electric and magnetic fields. These hyperfine interactions [22] gives rise to isomer shift (IS), the quadrupole splitting (QS) and the magnetic Zeeman splitting. Hence, in the present work, we put our efforts in exploring the magnetic and local interactions in $HoFe_{1-x}Cr_xO_3$ ($0 \leq x \leq 1$) compounds through magnetization measurements and Mössbauer spectroscopy.

**Experimental Details**:

Polycrystalline $HoFe_{1-x}Cr_xO_3$ ($0 \leq x \leq 1$) samples were prepared by conventional solid state reaction method using High purity oxide powders of $Ho_2O_3$, $Fe_2O_3$, and $Cr_2O_3$ (purity > 99.9%) (Sigma-Aldrich chemicals India) as starting raw materials. The temperature (T) dependent magnetization (M) (M *vs.* T) measurements were performed using a Quantum Design magnetic property measurement system (MPMS) in the temperature range of 5 – 300 K. M *vs.* T measurements were performed in zero field cooling (ZFC) as well as field cooling (FC) conditions. Essentially, in ZFC conditions, the compound was cooled from 300 K to 5 K without application of any magnetic field. However, in case of FC condition, the compound was cooled from 300 K in the presence of 1000 Oe up to a desired temperature, 5 K. Magnetization measurements in the temperature range of 300 – 900 K were carried out using a high temperature vibrating sample magnetometer (HT-VSM) (LakeShore make). The $^{57}$Fe Mössbauer measurements were carried out in transmission mode with a $^{57}$Co (Rh) radioactive source in constant acceleration mode using a standard PC-based Mössbauer spectrometer equipped with a WissEl velocity drive.

**Results and discussion**

The phase purity of all the samples HoFe$_{1-x}$Cr$_x$O$_3$ ($0 \leq x \leq 1$) have been confirmed [24] at room temperature using a powder x - ray diffraction (XRD) (PANalytical) with Cu - K$_\alpha$ radiation ($\lambda$ = 1.5406 Å) and with a step size of $0.017^0$ in the wide range of the Bragg's angle 2θ ($20^0$ - $80^0$) as shown in Fig.1. From the Fig.1, it is evident that the indexed reflections are allowed for a compound with GdFeO$_3$ type distorted perovskite structure with a space group *Pbnm*. In addition to the parent phase HoFeO$_3$ for $x$ = 0 compound, XRD infer that indeed there exists a small peak corresponding to iron garnet (Ho$_3$Fe$_5$O$_{12}$) phase at around 32° and 35° [24-25]. We do not see any impurity phase in case of other compounds with $x$ = 0.25, 0.5, 0.75 and 1 within the detectable limits of the XRD. It is worth noting that the rare-earth iron garnet (R$_3$Fe$_5$O$_{12}$) phase is stable compared to orthoferrites (RFeO$_3$) at high temperature [26]. The percentage of observed garnet phase, Ho$_3$Fe$_5$O$_{12}$ is around 5-6 %. The structural parameters obtained from the Rietveld refinement has been published elsewhere [24]. The decrease in the lattice parameter is evident from the refinement with increasing Cr$^{3+}$ content, which is consistent with the fact that ionic radius of $Fe^{3+}$ (0.645 Å) is larger than that of $Cr^{3+}$ (0.615 Å).

Now we shall discuss magnetization behaviour of $HoFe_{1-x}Cr_xO_3$ ($0 \leq x \leq 1$) compounds at high temperatures and subsequently explore the low temperature magnetization behaviour. Fig. 2(a) and 2(b) depicts the ZFC curves pertinent to susceptibility $\chi$ of the $HoFe_{1-x}Cr_xO_3$ ($0 \leq x \leq 1$) compounds at high temperature (300-900 K) and at low temperature (5-300 K) respectively with a magnetic field strength of 1000 Oe. From $\chi$ vs. T graph (Fig. 2) it is evident that indeed there exist various transitions pertinent to antiferromagnetic coupling (AFM), spin re-orientation transition (SR) and $Ho^{3+}$ ordering. However, above transitions exists at various temperature ranges. Below we discuss physical significance, variation and origin of such transitions in a detailed way.

Initially we discuss about the evidenced antiferromagnetic transition and its variation in $HoFe_{1-x}Cr_xO_3$ ($0 \leq x \leq 1$) compounds with respect to temperature. From Fig. 2 it is evident that a sudden jump prevails at 647 K for $x = 0$ compound, which can be attributed to the antiferromagnetic ordering of $Fe^{3+}$ moments in $HoFeO_3$ compound [14]. The increase in the magnetization below this transition indicates a weak ferromagnetic moment (WFM) arised due to a canted nature of $Fe^{3+}$ moments. Apart from the above transition, another transition is evident for $x = 0$ compound at around 550 K, which corresponds to $Fe^{3+}$ ordering in the rare earth iron garnet phase, $Ho_3Fe_5O_{12}$ [25]. We do not see any transition pertinent to iron garnet phase in the compounds other than parent compound. With $Cr^{3+}$, the Néel temperature ($T_N$) is varied in the range of 647 K ($x = 0$) to 142 K ($x = 1$) for the compounds $HoFe_{1-x}Cr_xO_3$ ($0 \leq x \leq 1$). Fig. 3 depicts the variation of $T_N$ pertinent to $HoFe_{1-x}Cr_xO_3$ compounds which were obtained from $\chi$ vs. T graphs. Essentially $\chi$ vs. T graphs demonstrated an antiferromagnetic transition at 647 K, 495 K, 273 K, 157 K, and 142 K for $x = 0$, 0.25, 0.50, 0.75 and 1.0 compounds respectively. Indeed, this indicates a decrease in the value of $T_N$ with $Cr^{3+}$ addition. Such a variation of $T_N$ with $Cr^{3+}$ could be due to the larger ionic radius of $Fe^{3+}$ (0.645 Å) in comparison with $Cr^{3+}$ (0.615 Å). In our work, since we substituted $Cr^{3+}$ to $Fe^{3+}$, there is a contraction of the lattice and hence the distortion in the crystal structure. Variation of lattice parameter with $Cr^{3+}$ addition has been reported elsewhere [24]. It has been reported that the structural distortion is closely related to the superexchange interaction and may influence $T_N$ in orthorhombic rare-earth ortho-ferrites ($RFeO_3$) [27] and ortho-chromites ($RCrO_3$) [28]. According to Goodenough-Kanamori rule [17], two adjacent transition metal ions interact through the superexchange interaction via a virtual charge transfer. The outer shell electron configuration for $Fe^{3+}$ and $Cr^{3+}$ are $t^3e^2$ and $t^3e^0$ respectively. In case of $RFeO_3$, the π bonding of half-filled $t^3$-O-$t^3$ orbitals and σ bonding of half-filled $e^2$-O-$e^2$ orbitals interact through superexchange interaction and are antiferromagnetic (AFM) in nature, which satisfies the Hund's rule. From this it can inferred that the hybridization of $t - e$ orbitals doesn't give any extra magnetic phase component to the overall superexchange interaction. On the other hand, in $RCrO_3$, $t - e$ hybridization gives a ferromagnetic (FM) component in addition to the existing AFM interaction in $t^3$-O-$t^3$ orbitals though the superexchange interaction [28]. As there exists both FM and AFM competing interactions through $t - e$ hybridization in Cr rich compounds, by the substitution of $Cr^{3+}$ to $Fe^{3+}$, the strength of FM interaction increases and AFM interaction diminishes, which may results in decrease of $T_N$. From the above, we believe that the decrease in $T_N$ with $Cr^{3+}$ is not only due to the weakening of Fe(Cr)-O-Fe(Cr) AFM exchange interaction, but also due to the enhancement in FM interaction between adjacent $Cr^{3+}$ moments through $t - e$ hybridization as a result of the structural distortion. In addition, molecular field theory is also used to quantify the values of $T_N$ up on $Cr^{3+}$ addition using the following relation. Essentially the relation between $J$ and $T_N$ can be defined as [29]

$$J/k = \frac{3T_N}{2zS(S+1)} \quad \ldots (1)$$

Where '$z$' represents the number of nearest neighbours and its value is 6 for ortho-ferrites, '$k$' denotes the Boltzmann's constant ($8.617 \times 10^{-5}$ eV/K), $S = 5/2$ and $3/2$ are the values pertinent to the spins of $Fe^{3+}$ and $Cr^{3+}$ ions respectively. Physically, $J$ signifies the strength of the exchange interaction between a pair of nearest magnetic ions in the ortho-ferrites. Estimated values of $J$ using equation (1) for $HoFeO_3$ and $HoCrO_3$ compounds are 18.4 K and 9.4 K respectively, which is in striking coincidence with the higher value of $T_N$ for $HoFeO_3$ (647 K) in comparison with the value pertinent to $HoCrO_3$ (142 K).

Now we discuss about shift in SR transition with $Cr^{3+}$ in $HoFe_{1-x}Cr_xO_3$ ($0 \leq x \leq 1$) compounds. From Fig. 2(b) it is evident that apart from a prominent peak pertinent to $T_N$, there exists a peak which corresponds to SR transition ($T_{SR}$) for the compounds with composition $x = 0$ ($T_{SR} = 30$ K and 50 K), $x = 0.25$ ($T_{SR} = 75$ K) and $x = 0.5$ ($T_{SR} = 150$ K) [30]. We ascribe such $T_{SR}$ in $HoFe_{1-x}Cr_xO_3$ ($0 \leq x \leq 1$) compounds to complex exchange interaction between the $Fe^{3+}$ and the $Ho^{3+}$ ions [31]. For instance, in case of $x = 0$ compound, multiple SR transitions are evident at 50 K and 30 K, the former one at around 50 K is a typical transition in rare earth ortho-ferrites and is consistent with literature [30]. The latter one around 30 K may be due to competing Zeeman and Van Vleck mechanism [32], which vanishes up on replacing the $Fe^{3+}$ ions with the $Cr^{3+}$ ions. From Fig. 2(b) it is clear that the $\chi$ decreases below the $T_{SR}$ for all the compositions ($x = 0$, 0.25 and 0.5), which demonstrates the rotation of $Fe^{3+}$ moments from one crystal axis to another. The shift in the $T_{SR}$ towards higher temperatures for the $x = 0.25$ and 0.5 compound may be ascribed to complex exchange interaction between the $Fe^{3+}/Cr^{3+}$ and the $Ho^{3+}$ ions [31]. The $T_{SR}$ takes place at 50 K, for $x = 0$ compound due to the domination of $Ho^{3+}$- $Fe^{3+}$ interaction over the $Fe^{3+}$- $Fe^{3+}$ interaction. As we substituted the $Cr^{3+}$ ion at the $Fe^{3+}$ site, this may weakens the $Fe^{3+}$-$Fe^{3+}$ interaction and may demand a shift in the $T_{SR}$ to a higher temperature. The possible reason can be explained as follows. In HoFeO$_3$, the energy levels related with Ho$^{3+}$ ions lie in the orthorhombic crystal field and molecular field formed by the sublattices of Fe$^{3+}$ and Ho$^{3+}$. The degeneracy of the ground state multiplet $^5I_8$ of free non-Kramer Ho$^{3+}$ ion essentially be lifted by the crystalline field and gives various singlets states. In particular, it has been observed that the ground state singlets (accidental doublet) are isolated from the other crystal field splitting states with an energy difference of more than 80 cm$^{-1}$[33]. Highly anisotropic nature of *g*-factor pertinent to accidental doublet states would plays a predominant role in the magnetism of Ho$^{3+}$ in HoFeO$_3$ and the same has been well demonstrated by Griffith [34]. It also has been established that at low temperatures in case of HoFeO$_3$, aforementioned anisotropic nature of *g* – factor leads to an 'Ising' axis for the Ho$^{3+}$ moments in the *ab*-plane at an angle ± 63° with respect to the orthorhombic *a*-axis [35]. Thus a magnetic field parallel to *ab* - plane will split (Zeeman splitting) the accidental doublet and increases the energy of the system. Hence, the origin of SR transition in HoFeO$_3$ can be discussed on the basis of minimum energy configuration of spins which involve in complex magnetic interactions. Essentially, SR transition occurs due to the transformation of the high temperature $\Gamma_4$ (G$_x$A$_y$F$_z$) phase [high energy due to Zeeman splitting of accidental doublet] to $\Gamma_2$ (F$_x$C$_y$G$_z$) phase [low energy due to rotation of Fe$^{3+}$ moments towards c – axis by 90°] at low temperatures. As a result of $\Gamma_4 \rightarrow \Gamma_2$ phase transition, we do see a SR transition in HoFeO$_3$ at around 50 K. On the other hand, the SR transition strongly depends on the exchange interaction strength of Ho-Fe(Cr) and Fe(Cr)-Fe(Cr). We do see an increase in $T_{SR}$ with Cr$^{3+}$ composition. We believe that as we substitute Cr$^{3+}$ to Fe$^{3+}$, Fe – Fe exchange interaction diminishes and Ho – Fe interaction enhances. From the above explanation it is clear that due to an enhanced Ho – Fe interaction there is an enhanced $T_{SR}$ with Cr$^{3+}$ addition.

In addition to the AFM and SR transitions, we also could see a transition at low temperatures for $x = 0.25$ (8.5 K), $x = 0.5$ (13 K) and $x = 1$ (8 K) compounds, which can be attributed to $Ho^{3+}$ ordering [Fig. 2(b)] [25]. $\chi$ behaviour for x = 0.75 is quite different and the "diamagnetism" like negative magnetization is observed. Earlier, negative magnetization has been observed in the $HoCrO_3$ compound for a cooling field of 100 Oe, however, such behaviour is absent for 1000 Oe. This intriguing phenomenon has been explained on the basis of competing WFM-AFM interactions [36]. In contrast, negative magnetization has been reported in ferro/ferri/antiferromagnetic materials and such fascinating phenomena has been explained on the basis of small negative trapping fields in the sample space [37]. Nevertheless, neutron diffraction data on $HoFe_{1-x}Cr_xO_3$ solid solutions reravals no signature of negative magnetization [38]. Hence, we believe that the observed negative magnetization for x = 0.75 compound could be due to trapped fields.

To investigate the paramagnetic behaviour of $HoFe_{1-x}Cr_xO_3$ compounds, we have plotted $1/\chi$ vs. T from the ZFC magnetization curves. Fig. 4 depicts the temperature dependence of inverse susceptibility obtained for $HoFe_{1-x}Cr_xO_3$ ($0 \leq x \leq 1$) compounds. From Fig. 4, almost the linear behaviour is evident for the inverse molar magnetic susceptibility above the $T_N$. The total effective magnetic moment $\mu_{eff}$ of $HoFe_{1-x}Cr_xO_3$ compounds is obtained by fitting the paramagnetic region of $\chi^{-1}$ vs. T to Curie-Weiss law:

$$\chi = \frac{C}{T - \theta}$$

Where $C = \frac{N_A \mu_{eff}^2}{3k_B}$ is Curie constant, $N_A$ is the Avogadro's number, $\mu_B$ is the Bohr magneton, $\mu_{eff}$ is the effective magnetic moment, $k_B$ is the Boltzmann constant and $\theta$ is the paramagnetic Weiss temperature. After fitting the linear region above $T_N$, from the slope $\mu_{eff}$ is obtained and it is found to be 16.59 $\mu_B$, 11.80 $\mu_B$, 11.62 $\mu_B$, 11.15$\mu_B$, 11.13 $\mu_B$ for $x = 0$, 0.25, 0.5, 0.75 and 1 compounds respectively. The large value of $\mu_{eff}$ = 16.59 $\mu_B$ for $x = 0$, when compared to its theoretical value 12.14 $\mu_B$ can be ascribed to the impurity iron garnet $Ho_3Fe_5O_{12}$ phase formed along with the parent phase $HoFeO_3$. Except for the compound with $x = 0$, the estimated value of the $\mu_{eff}$ is found to be close to the theoretical values 11.93 $\mu_B$ ($x = 0.25$), 11.72 $\mu_B$ ($x = 0.5$), 11.50 $\mu_B$ ($x = 0.75$), 11.28 $\mu_B$ ($x = 1$) calculated from the free ion values of $Ho^{3+}$ (10.60 $\mu_B$), $Fe^{3+}$ (5.92 $\mu_B$) and $Cr^{3+}$ (3.87 $\mu_B$) (spin only values of $Fe^{3+}$ and $Cr^{3+}$) using the relation [39] $\mu_{theory} = \left[\mu_{Ho^{3+}}^2 + (1-x)\mu_{Fe^{3+}}^2 + x\mu_{Cr^{3+}}^2\right]^{1/2}$ and by assuming their randomness in the paramagnetic phase. The equation $\mu_{Fe^{3+}/Cr^{3+}} = 2[S(S+1)]^{1/2}$ is used to calculate $\mu_{Fe^{3+}}/\mu_{Cr^{3+}}$ moments and the formula $\mu_{Ho^{3+}} = |g_J|[J(J+1)]^{1/2}$ is used to calculate $\mu_{Ho^{3+}}$ moments. In the above equations $S$ is the spin sate of $\mu_{Fe^{3+}}/\mu_{Cr^{3+}}$ and $J$ represents the total angular momentum and $g_J$ represents the Landè $g$-factor respectively. For all the compounds, the paramagnetic Curie temperature is estimated by extrapolating the high temperature linear region of $1/\chi$ vs. T graph to x – axis and it is found to be negative as shown in Table 1. According to Goodenough-Kanamori rule [17], in order to observe ferromagnetic ground state in $HoFe_{1-x}Cr_xO3$ compounds, interaction should be of type $Fe^{3+}$ ($d^5$) – O – $Cr^{3+}$ ($d^3$) – O – $Fe^{3+}$ ($d^5$). However, from $1/\chi$ vs. T graph, we do observe negative values for $\theta$, which suggests anti-ferromagnetic ground state. Hence, we believe that $Fe^{3+}$ - O – $Fe^{3+}$, $Cr^{3+}$ - O – $Cr^{3+}$ and $Fe^{3+}$ - O – $Cr^{3+}$ superexchange interactions are present at B – site [40-41].

It has been reported that the magnetic properties of ortho-ferrites depends upon the oxidation state as well as the spin configuration of the transition metal cations [42]. Brinks *et al.* [43] observed a decrease in the $T_N$ in $Pr_{1-x}Sr_xFeO_3$ compounds due to an enhancement in the oxidation state of Fe from $Fe^{3+}$ to $Fe^{4+}$. In our present investigation, $T_N$ decreases with $Cr^{3+}$ in

HoFe$_{1-x}$Cr$_x$O$_3$ compounds from 647 K ($x = 0$) to 142 K ($x = 1$). In general, the variation of T$_N$ could be due to *(a)* structural distortion *(b)* change in oxidation state. From our phase purity investigation, structural distortions have been realized in HoFe$_{1-x}$Cr$_x$O$_3$ compounds. However, the oxidation of state of Fe in the present compounds is ambiguous. In this scenario, it is essential to know the knowledge of oxidation state of the transition metal, and its implications on the magnetic properties of HoFe$_{1-x}$Cr$_x$O$_3$ compounds, and hence, we performed $^{57}$Fe Mössbauer spectroscopy measurements. Fig. 5 depicts the room temperature $^{57}$Fe Mössbauer measurements carried out on polycrystalline HoFe$_{1-x}$Cr$_x$O$_3$ ($0 \leq x \leq 1$) compounds. Experimentally recorded data is represented by the plus "+" symbol, while the least square fit of the spectrum is represented by a solid line. Fitting results indicates a single sextet for $x = 0$ compound while doublet is evident for $x = 0.5$ and 0.75 compounds respectively. Since the hyperfine field of the garnet Ho$_3$Fe$_5$O$_{12}$ [44] and the parent compound HoFeO$_3$ are very close, we were not able to fit the Mössbauer data with garnet phase which is only about 5-6%. On the other hand, distribution of fields matches reasonably well with experimental data points in the case of $x = 0.25$ compound with an average hyperfine field value of 415.3 kOe. The fitted parameters are listed in the Table 2. Oxidation state of '+3' for Fe in HoFeO$_3$ is confirmed by realizing the six line spectra and corresponding values pertinent to hyperfine field (H$_{hf}$), isomer shift (IS) and quadrupole splitting (QS) are 497.6 kOe, 0.373 mm/s and 0.015 mm/s respectively. Indeed, observed H$_f$, IS and QS are within the range for an element to have '+3' oxidation state and is in accordance with the literature [45]. Since there is no change in oxidation state, we believe that variation of T$_N$ in HoFe$_{1-x}$Cr$_x$O$_3$ ($0 \leq x \leq 1$) compounds is due to structural distortion.

In general H$_{hf}$, IS and QS are very crucial and sensitive for a change in local environment of Fe$^{3+}$ particularly in ortho-ferrites. In the present work, we modified the local environment of Fe$^{3+}$ by substituting Cr$^{3+}$. In subsequent section we discuss about the variation of H$_{hf}$, IS and

QS parameters and related physical mechanism with $Cr^{3+}$ addition in $HoFe_{1-x}Cr_xO_3$ compounds at 300 K.

From Table – 2 it is evident that IS shift decreases with $Cr^{3+}$. The decrease in IS shift is due to decrease in radial part of $3d$ wave function with Cr substitution [46]. On the other hand, quadrupole splitting is a parameter which gives the information about the lifting of the degeneracy of a nuclear state with $I > ½$ by experiencing the electric field gradient (EFG) due to its interaction with an asymmetric electronic environment surrounding to it. The small value of QS ~ 0.015 mm/s observed for $HoFeO_3$ suggests that Fe is almost in octahedral environment (cubic electric field gradient) [22]. The value of QS increases with an increase in Cr compostion, and is quite appreciable for $x \geq 0.5$ compounds. This increase in QS may be ascribed to the distortion in $FeO_6$ octahedra by substituting $Cr^{3+}$ ions at Fe site [24], due to which the Fe atom experience a non cubic electric field and which leads to increase in QS.

The six resonance lines observed in Mössbauer spectrum for $HoFeO_3$ at room temperature indicates that the $Fe^{3+}$ ions experience an internal magnetic field. It has been reported that $HoFeO_3$ is antiferromagnetically ordered below 641 K [14]. In ortho-ferrites, the antiferromagnetically aligned iron ions with small canting angle with respect to the AFM axis results a small net ferromagnetic moment. The value of the calculated hyperfine field ($H_{hf}$) for $HoFeO_3$ at room temperature is 497.6 kOe (typical value for $Fe^{3+}$ ion), which is in agreement with previously reported value [45]. With an increase in the concentration of Cr, the strength of AFM interaction decreases, resulting in a decrease in the internal magnetic field experienced by the $Fe^{3+}$ ion at 300 K. As the compounds with composition $x \geq 0.5$ consists ordering temperatures below 300 K, disappearance of six line spectra is evident which may be due to vanishing of internal magnetic fields at 300 K.

**Conclusions**:

In summary, we have explored the magnetic and Mössbauer properties of polycrystalline perovskite HoFe$_{1-x}$Cr$_x$O$_3$ ($0 \leq x \leq 1$) compounds. A perfect correlation between the results of structural, Mössbauer and magnetic properties are obtained. The observed decrease of $T_N$ and increase of $T_{SR}$ with Cr doping was attributed to the dilution of Ho$^{3+}$-Fe$^{3+}$ and Fe$^{3+}$-Fe$^{3+}$ interactions in the entire temperature region and the enhancement of ferromagnetic interaction of adjacent Cr$^{3+}$ moments through $t$-$e$ hybridization as a result of the structural distortion. The observed decrease in IS with Cr content infers the increase the interaction between the effective nuclear charge with the 3$s$ electrons. The small QS value observed for HoFeO$_3$ indicates the local environment remains octahedral at the unit cell. Increase in a value of QS with Cr doping is due to the non-cubical electric field gradient acting on Fe atom arised from the distortion in FeO$_6$ octahedra by substituting Cr$^{3+}$ ions at Fe site. The decrease in hyperfine field was ascribed to a decrease in Fe$^{3+}$-Fe$^{3+}$ AFM interaction with the addition of Cr. The disappearance of six line Mössbauer spectra for x ≥ 0.5 infers the compounds are in paramagnetic in nature at 300 K.

**Acknowledgements**

We would like to acknowledge Indian Institute of Technology, Hyderabad and Department of Science and Technology (DST) (Project #SR/FTP/PS-190/2012) for the financial support. We are grateful to Dr. Alok Banerjee and Mr. Kranthi Kumar, UGC-CSR, Indore for carrying out the magnetization measurements. We also grateful to UGC-DAE Consortium (Project # CSR-IC/CRS-162/2015-16/19) for the financial support and their extended facilities.

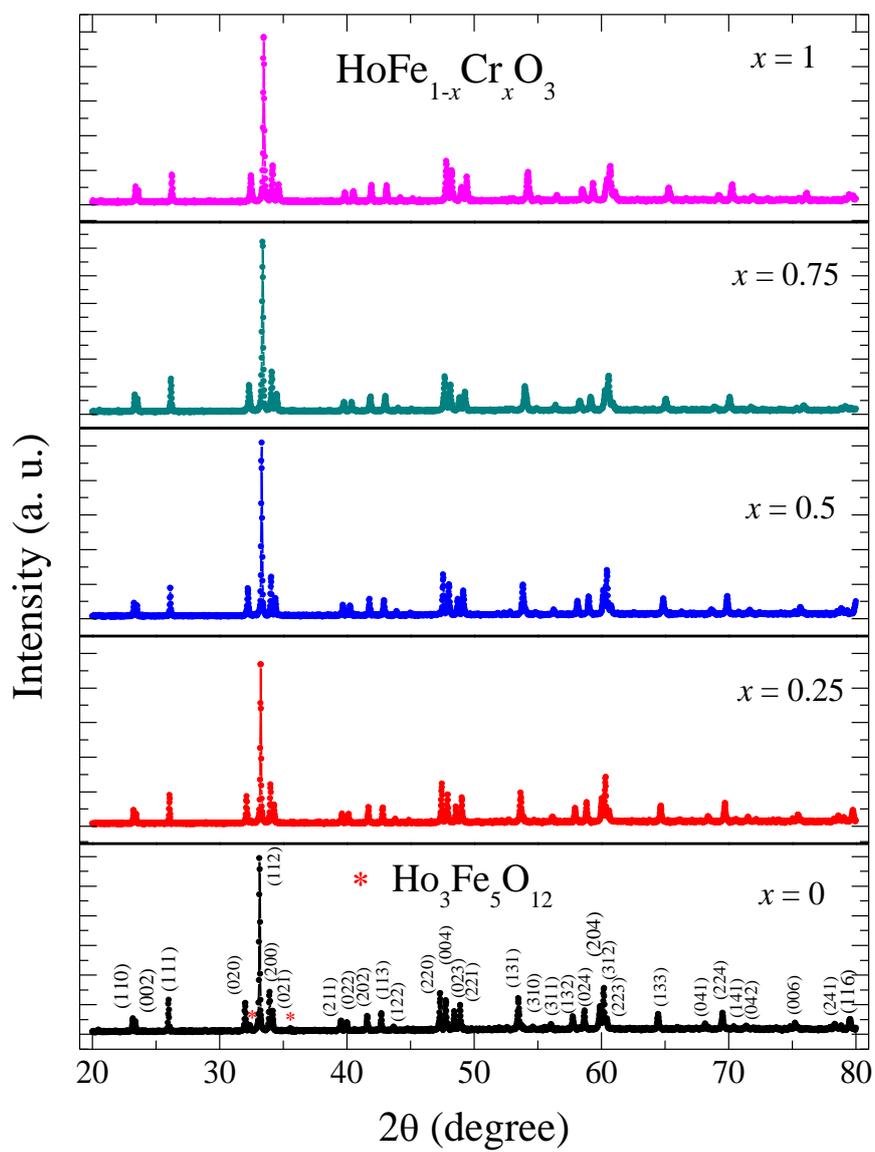

**Fig. 1:** Powder x-ray diffraction patterns HoFe$_{1-x}$Cr$_x$O$_3$ ($0 \leq x \leq 1$) compounds recorded at room temperature.

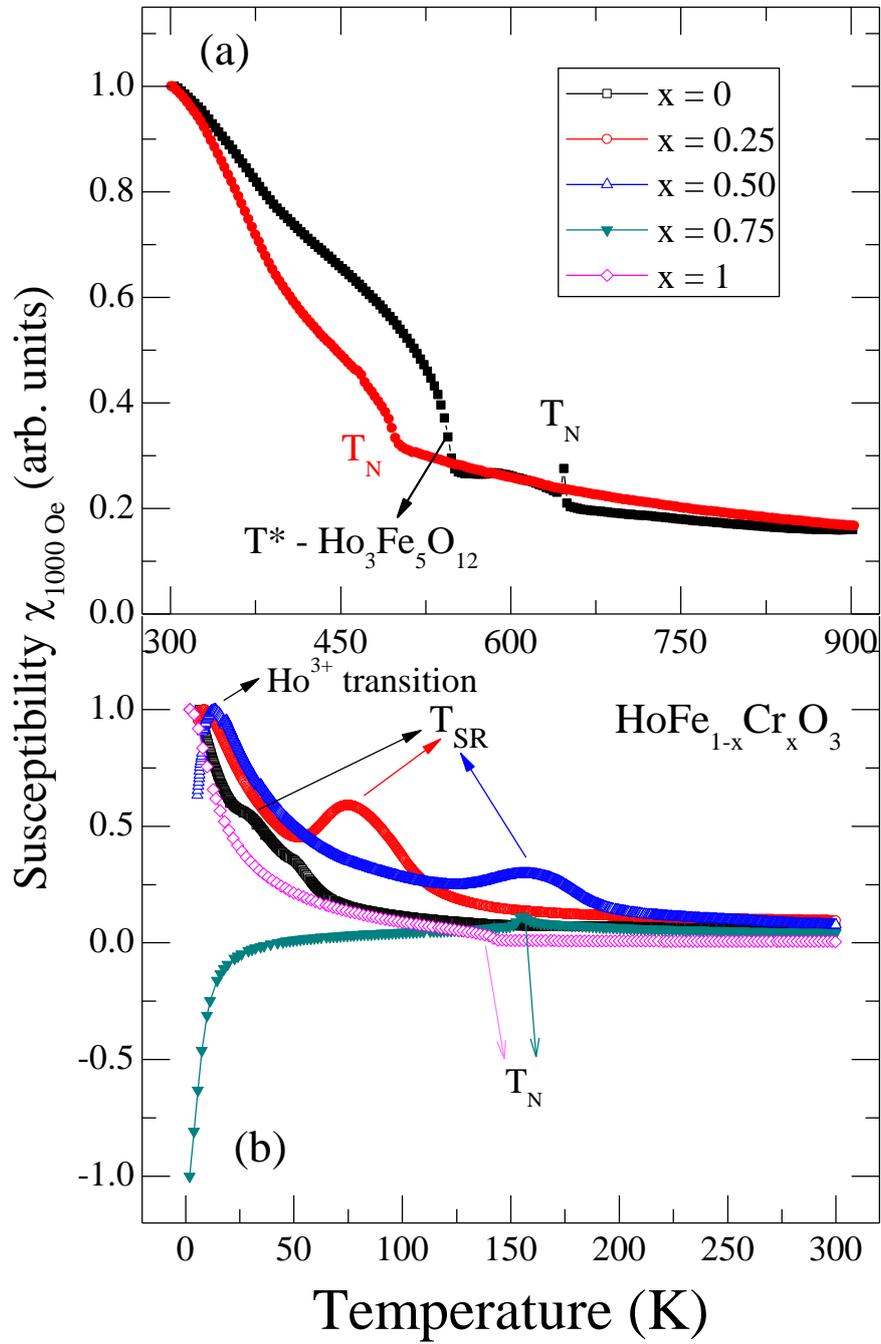

**Fig.2**: Temperature (T) dependence of susceptibility (χ) pertinent to normalized magnetization (M) in (a) high temperature and (b) low temperature region for $HoFe_{1-x}Cr_xO_3$ ($0 \leq x \leq 1$) compounds.

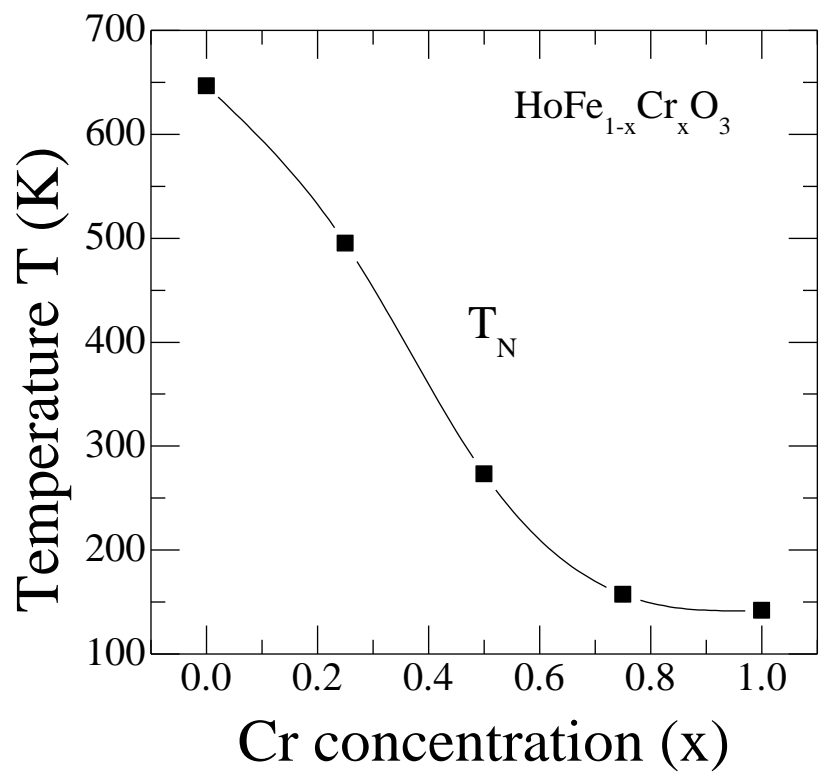

**Fig. 3**: Variation of Néel temperature $T_N$ of $HoFe_{1-x}Cr_xO_3$ ($0 \leq x \leq 1$) compounds with $Cr^{3+}$ concentration.

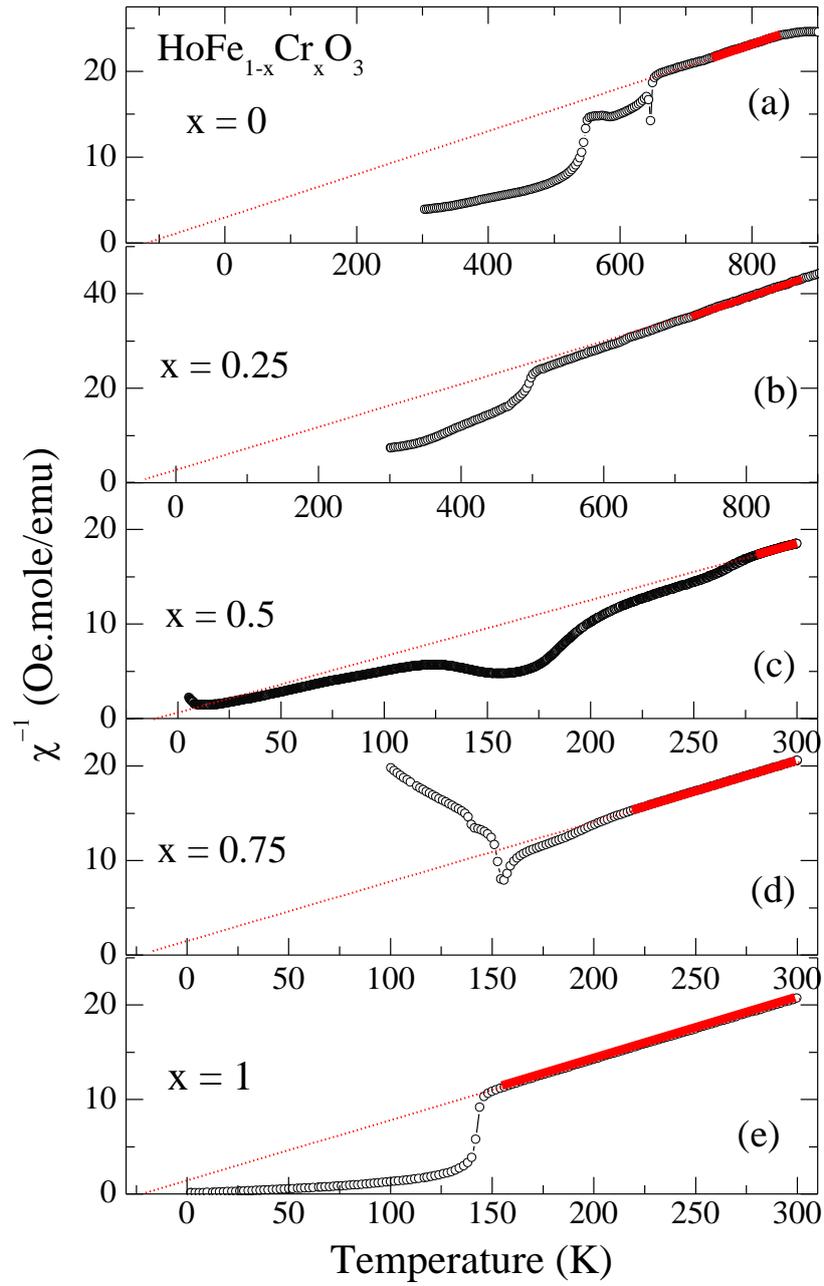

**Fig. 4**: Temperature dependence of inverse molar susceptibility of HoFe$_{1-x}$Cr$_x$O$_3$ ($0 \leq x \leq 1$) compounds measured at 1000 Oe in the range of 5 – 300 K. Small dots represents the best fit of Curie-Weiss law.

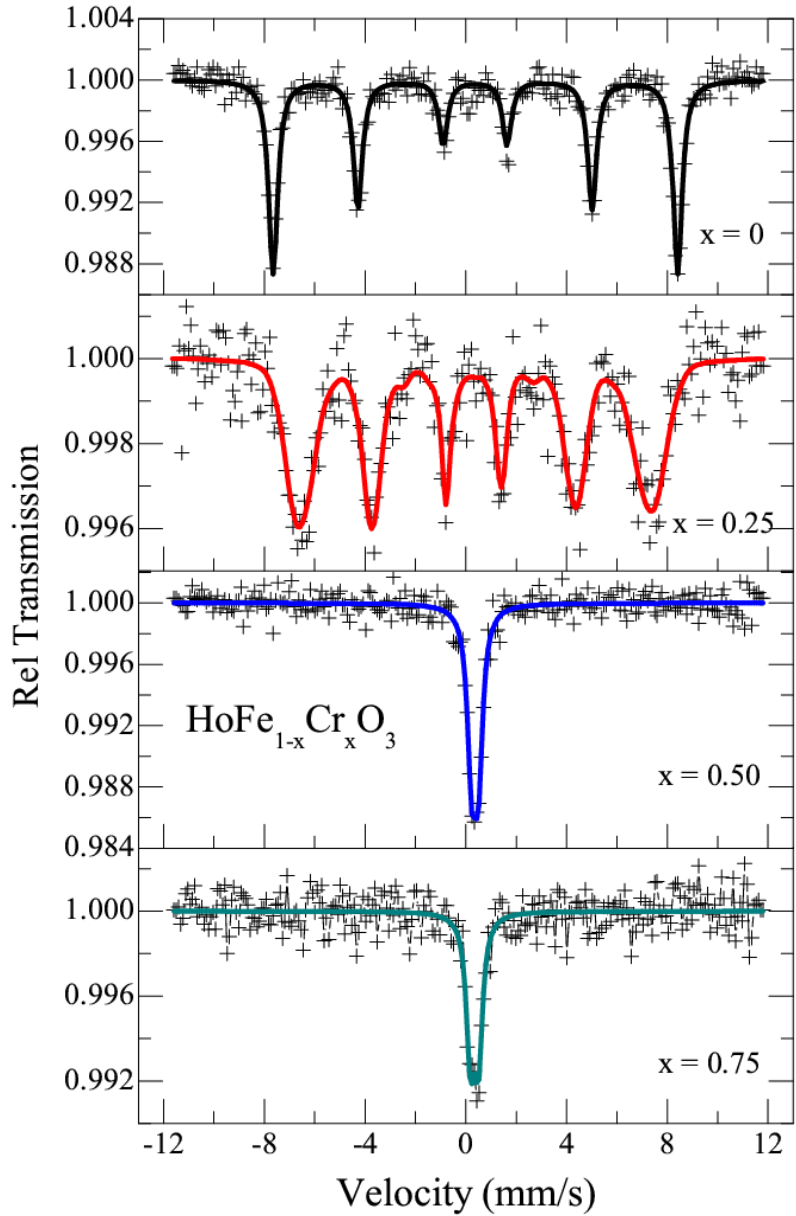

**Fig. 5**: Room temperature $^{57}$Fe Mössbauer spectrum of polycrystalline HoFe$_{1-x}$Cr$_x$O$_3$ ($0 \leq x \leq 1$) compounds. The experimental data is represented by the plus "+" symbol and the solid line is the best fit of the data.

**Table. 1**: Effective magnetic moment $\mu_{eff}$, paramagnetic Curie temperature $\theta$ of $HoFe_{1-x}Cr_xO_3$ ($0 \leq x \leq 1$) compounds obtained from the best fit of $\chi^{-1}$ vs. T by Curie-Weiss law.

| Compound $HoFe_{1-x}Cr_xO_3$ | $\mu_{eff}$ (experimental) | $M_{the}$ (theoretical) | Paramagnetic curie temperature ($\theta$) |
|---|---|---|---|
| $x = 0$ | 16.59 $\mu_B$ | 12.14 $\mu_B$ | -118.5 K |
| $x = 0.25$ | 11.80 $\mu_B$ | 11.93 $\mu_B$ | -41 K |
| $x = 0.5$ | 11.62 $\mu_B$ | 11.72 $\mu_B$ | -10 K |
| $x = 0.75$ | 11.15 $\mu_B$ | 11.50 $\mu_B$ | -16.84 K |
| $x = 1$ | 11.13 $\mu_B$ | 11.28 $\mu_B$ | -19 K |

**Table 2**: Parameters obtained from Mössbauer spectroscopy for $HoFe_{1-x}Cr_xO_3$ ($0 \leq x \leq 1$) compounds at 300 K.

| | FWHM (mm/s) | IS (mm/s) | QS (mm/s) | $H_{hf}$ (kOe) |
|---|---|---|---|---|
| $x = 0$ | 0.430 ± 0.018 | 0.373 ± 0.005 | 0.015 ± 0.010 | 497.6 ± 0.4 |
| $x = 0.25$ | 0.45 | -- | -- | 415.3 (Avg) |
| $x = 0.50$ | 0.408 ± 0.019 | 0.368 ± 0.009 | 0.26 ± 0.01 | --- |
| $x = 0.75$ | 0.388 ± 0.049 | 0.342 ± 0.018 | 0.32 ± 0.03 | --- |